\documentclass[12pt,a4paper,twoside]{article}
\newenvironment{changemargin}[2]{%
\begin{list}{}{%
\setlength{\leftmargin}{#1}%
\setlength{\rightmargin}{#2}%
}%
\item[]}
{\end{list}}
\usepackage{graphicx}
\usepackage{color}
\usepackage{lscape}
\usepackage{hyperref}
\usepackage{epstopdf}
\usepackage{lscape}
\usepackage{amsmath}
\usepackage{hyperref}
\usepackage{amsmath}
\usepackage{setspace}
\begin{document}
\baselineskip=0.20in
{\bf \LARGE
\begin{changemargin}{-1.2cm}{0.5cm}
\begin{center}
Effect of oblateness, radiation and a circular cluster of material points on the  stability of equilibrium points in the restricted four-body problem
\end{center}
\end{changemargin}}
\vspace{4mm}
\begin{center}
{\Large{\bf Babatunde J. Falaye $^\dag$}}\footnote{\scriptsize E-mail:~ fbjames11@physicist.net;~ babatunde.falaye@fulafia.edu.ng}
\end{center}
{\small
\begin{center}
{\it Applied Theoretical Physics Division, Department of Physics, Federal University Lafia,  P. M. B. 146, Lafia, Nigeria.}
\end{center}}
\vspace{4mm}
\begin{center}
Few Body Sys. (2014) DOI: 10.1007/s00601-014-0922-3
\end{center}
\vspace{4mm}
\begin{abstract}
\noindent
Within the framework of restricted four-body problem, we study the motion of an infinitesimal mass by assuming that the primaries of the system are radiating-oblate spheroids surrounded by a circular cluster of material points. In our model, we assume that the two masses of the primaries $m_2$ and $m_3$ are equal to $\mu$ and the mass $m_1$
is $1-2\mu$. By using numerical approach, we have obtained the equilibrium points and examined their linear stability. The effect of potential created by the circular cluster and oblateness coefficients for the more massive primary and the less massive primary, on the existence and linear stability of the libration point have been critically examine via numerical computation.  The stability of these points examined shows that the collinear and the non-collinear equilibrium points are unstable. The result presented in this paper have practical application in astrophysics.
\end{abstract}


\section{Introduction}
In recent years, there has been a growing interest in studying three-body problem with the aim of approximating the behavior of real celestial systems (\cite{N1,A2} and refs. therein). The results from the study have been found useful in the theory of dynamical systems and in astronomy. The simplest form of the three-body problem is restricted three-body problem (R3BP). It describe the motion of an infinitesimal mass moving under the gravitational influence of two massive bodies called the primaries which moves in circular orbits around their center of mass on account of their mutual attraction and the infinitesimal mass not influencing the motion of the primaries. The classical restricted three-body problem possesses five equilibrium points. The first three points $(L_{1}, L_2, L_3)$, denotes the collinear points while the last two points ($L_{4}$ and $L_5$) denotes the triangular points. The Earth-Moon system together with an artificial satellite constitute such problem \cite{N1}.

Now, with an interest in four-body problem. Couple R3BP is one of the examples of restricted four-body problem (R4BP). The smaller body is called as the infinitesimal mas while the remaining massive bodies are called as the primaries. very recently, Batagiannis and Papadakis \cite{A1} studied the families of simple symmetric and non-symmetric periodic orbits in the restricted four body problem. Few among other interesting works are Kumari and Kushvah \cite{N2}, Papadouris and Papadakis \cite{N3}, Alvarez and Vidal \cite{N4}. The purpose for studying R4BP includes their application in general behavior of the synchronous orbit in presence of Moon and the Sun. 

In this paper, our aim is to study the motion of an infinitesimal mass by assuming that the primaries of the system are radiating-oblate spheroids surrounded by a circular cluster of material points. By using numerical computation procedure via Wolfram MATHEMATICA v10, we locate the libration points and examine their linear stability under the effect of oblateness and gravitational potential from a circular cluster of material points.

\section{Equation of motion}
Let $m_i$ $(i=1,2,3)$ be the masses of the primaries with $m_1>m2\geq m_3$ moving in circular periodic orbits around their center of mass fixed at the origin of the
coordinate system. These masses always lies at the vertices of equilateral triangle with $m_1$ being on the positive x-axis at the origin of time. The motion of the system is referred to axes rotating with uniform angular velocity \cite{A1}. The three bodies move in the same plane and their mutual distances remain unchanged with respect to time. The motion of the primaries consists of circular orbits around their center of gravity.  An equilateral equilibrium configuration of the three-bodies which is a particular solution of the three-body problem given by Lagrange is formed by the primaries at any instant of a given time. We assume that the influence of infinitesimal mass on the motion of primaries moving under their mutual gravitational attraction is negligible.  For studying the position of the infinitesimal mass, m, in the plane of motion of the primaries we applied the synodical coordinates. Thus the masses of the primaries are fixed at
\begin{eqnarray}
 \left(\sqrt{3}\mu,0\right),\ \ \ \left(-\frac{\sqrt{3}}{2}(1-2\mu), -\frac{1}{2}\right) \ \ \mbox{and} \ \ \left(-\frac{\sqrt{3}}{2}(1-2\mu),\frac{1}{2}\right)
\end{eqnarray}
where the mass parameter $\mu$ is taken as
\begin{equation}
\mu=\frac{m_2}{m_1+m_2+m_3}=\frac{m_3}{m_1+m_2+m_3}.
\end{equation}
We now take the potential energy of the infinitesimal mass, under the influence of a circular cluster of material points with its center at the origin of the coordinate system , under radiating-oblate primaries as
\begin{equation}
V=-Gm\left[q_1(1-2\mu)\left(\frac{1}{r_1}+\frac{A_1}{2r_1^2}-\frac{3A_2}{8r_1^5}\right)+q_2\mu \left(\frac{1}{r_2}+\frac{B_1}{2r_2^3}-\frac{3B_2}{8r_2^5}\right)+\frac{\mu}{r_3}+\frac{M_b}{\left(r^2+T^2\right)^{1/2}}\right].
\end{equation}
with $r_i(i=1,2,3)$ define as
\begin{eqnarray}
r_1=\sqrt{\left(x-\sqrt{3}\mu\right)^2+y^2},\ \ \ r_2=\sqrt{\left(x+\frac{\sqrt{3}}{2}(1-2\mu)\right)^2+\left(y-\frac{1}{2}\right)^2}\nonumber \\
r_3=\sqrt{\left(x+\frac{\sqrt{3}}{2}(1-2\mu)\right)^2+\left(y+\frac{1}{2}\right)^2}
\end{eqnarray}
where G is the gravitational constant. $q_1$ and $q_2$ are the radiation factors of the  primaries. $q_i = 1-\frac{F_{r_i}}{F_{g_i}}$ $(i=1,2)$, where $F_r$ is the force caused by radiation pressure force and $F_g$ results due to gravitational force \cite{A2,A3}. the term $\frac{M_b}{\left(r^2+T^2\right)^{1/2}}$ is the potential due to the circular cluster of material points (\cite{A3, A4} and refs therein) where $M_b$ is the total mass of the circular cluster of material points. the oblateness coefficients for the more
massive primary denoted as $A_i$ $\left(A_i = J_{2i}R_1^{2i}\right)$ and for the less massive primary as $B_i$  $\left(Bi = J_{2i}R^{2i}_2\right)$ with $J_{2i}$ being the zonal harmonic coefficients and $R_{1,2}$ denote the mean radii of $m_{1,2}$, supposing that the primaries have their equatorial planes coinciding with the plane of motion. Also, $r$ is the distance of the infinitesimal mass and which is given by
\begin{table}[!t]
{\scriptsize
\caption{\normalsize The collinear equilibrium points } \vspace*{10pt}{
\begin{tabular}{llcccccccc}\hline\hline\\[1ex]
\multicolumn{2}{l}{}&\multicolumn{3}{c}{$B_1=0.0015$, $B_2=0$, $M_b=0$ and $A_2=0$}&\multicolumn{2}{r}{}&\multicolumn{3}{r}{$B_1=0.0015$, $A_2=B_2=0.0001$, $M_b=0.01$}\\[1.5ex]
{}&{}&{}&{}&{}&{}&{}&{}&{}&{}\\[-0.5ex]\hline
$A_1$ &&&      $L_1$&      $L_2$&&&&      $L_1$&      $L_2$\\[1ex]\hline
0.0000&&&-0.95306838& 1.12278209&&&&-0.94898914& 1.11885937\\[1ex]
0.0015&&&-0.95221252& 1.12310149&&&&-0.94815485& 1.11919521\\[1ex]
0.0030&&&-0.95135906& 1.12341751&&&&-0.94732288& 1.11952752\\[1ex]
0.0045&&&-0.95050799& 1.12373023&&&&-0.94649324& 1.11985634\\[1ex]
0.0060&&&-0.94965932& 1.12403970&&&&-0.94566591& 1.12018174\\[1ex]
0.0075&&&-0.94881302& 1.12434596&&&&-0.94484088& 1.12050378\\[1ex]
0.0090&&&-0.94796909& 1.12464908&&&&-0.94401815& 1.12082252\\[1ex]
\hline
\multicolumn{2}{l}{}&\multicolumn{3}{c}{$A_1=0.0015$, $B_2=0$, $M_b=0$ and $A_2=0$}&\multicolumn{2}{r}{}&\multicolumn{3}{r}{$A_1=0.0015$, $A_2=B_2=0.0001$ and $M_b=0.01$}\\[1.5ex]
{}&{}&{}&{}&{}&{}&{}&{}&{}&{}\\[-0.5ex]\hline
$B_1$ &&&      $L_1$&      $L_2$&&&&      $L_1$&      $L_2$\\[1ex]\hline
0.0000&&&-0.95252453& 1.12376801&&&&-0.94844746& 1.11984694\\[1ex]
0.0015&&&-0.95221252& 1.12310149&&&&-0.94815485& 1.11919521\\[1ex]
0.0030&&&-0.95189701& 1.12243703&&&&-0.94785857& 1.11854546\\[1ex]
0.0045&&&-0.95157813& 1.12177461&&&&-0.94755869& 1.11789767\\[1ex]
0.0060&&&-0.95125599& 1.12111423&&&&-0.94725548& 1.11725183\\[1ex]
0.0075&&&-0.95093069& 1.12045588&&&&-0.94694891& 1.11660793\\[1ex]
0.0090&&&-0.95060234& 1.11979954&&&&-0.94663912& 1.11596596\\[1ex]	  
\hline
\end{tabular}\label{tab1}}
\vspace*{-1pt}}
\end{table}

\begin{landscape}
\begin{table}[!t]
{\tiny
\caption{\normalsize The non-collinear equilibrium points } \vspace*{10pt}{
\begin{tabular}{cccccccccc}\hline\hline\\[1ex]
\multicolumn{4}{l}{}&\multicolumn{2}{c}{$B_1=0.0015$}&\multicolumn{4}{c}{}\\[1.5ex]
{}&{}&{}&{}&{}&{}&{}&{}&{}&{}\\[-0.5ex]\hline
$A_1$ &&$L_3 (B_2=A2=M_b=0)$&$L_3 (B_2=A2=10^{-4}, M_b=0.01)$&&$L_4 (B_2=A2=M_b=0)$& $L_4 (B_2=A2=10^{-4}, M_b=0.01)$ &&$L_5 (B_2=A2=M_b=0)$& $L_5 (B_2=A2=10^{-4}, M_b=0.01)$\\[1ex]\hline
0.0000&&(-0.19345696,-0.28884586)&(-0.20144084,-0.28582247)&&(-0.87681266,-0.82897112)&(-0.87422161,-0.82679301)&&(-0.87791426,0.83013621)&(-0.87491830,0.82756065)\\[1ex]
0.0030&&(-0.19443441,-0.28989552)&(-0.20225222,-0.28692059)&&(-0.87599362,-0.82841022)&(-0.87342321,-0.82624651)&&(-0.87709873,0.82957741)&(-0.87412052,0.82701382)\\[1ex]
0.0060&&(-0.19539083,-0.29092004)&(-0.20304983,-0.28799216)&&(-0.87518047,-0.82785288)&(-0.87263035,-0.82570342)&&(-0.87628899,0.82902218)&(-0.87332828,0.82647040)\\[1ex]
0.0090&&(-0.19632718,-0.29192061)&(-0.20383411,-0.28903841)&&(-0.87437306,-0.82729906)&(-0.87184299,-0.82516371)&&(-0.87548498,0.82847047)&(-0.87254151,0.82593035)\\[1ex]\hline	
\multicolumn{4}{l}{}&\multicolumn{2}{c}{$B_1=0.0015$}&\multicolumn{4}{c}{}\\[1.5ex]
{}&{}&{}&{}&{}&{}&{}&{}&{}&{}\\[-0.5ex]\hline
$A_1$ &&$L_6 (B_2=A2=M_b=0)$& $L_7 (B_2=A2=10^{-4}, M_b=0.01)$ &&$L_8 (B_2=A2=M_b=0)$& $L_4 (B_2=A2=10^{-4}, M_b=0.01)$ &&$L_5 (B_2=A2=M_b=0)$& $L_5 (B_2=A2=10^{-4}, M_b=0.01)$\\[1ex]\hline
0.0000&&(-0.19197731,0.28831476)&(-0.20074188,0.28562894)&&(0.17004318,0.91238631)&(0.16893314,0.90967858)&&(0.16892433,-0.91225463)&(0.16796331,-0.90956875)\\[1ex]
0.0030&&(-0.19295597,0.28935976)&(-0.20155940,0.28672765)&&(0.16821900,0.91227333)&(0.16711754,0.90958927)&&(0.16709850,-0.91213880)&(0.16614685,-0.90947703)\\[1ex]
0.0060&&(-0.19391344,0.29037964)&(-0.20236301,0.28779979)&&(0.16640949,0.91215568)&(0.16531652,0.90949493)&&(0.16528733,-0.91201832)&(0.16434498,-0.90938031)\\[1ex]
0.0090&&(-0.19485068,0.29137562)&(-0.20315319,0.28884661)&&(0.16461444,0.91203350)&(0.16352989,0.90939570)&&(0.16349063,-0.91189334)&(0.16255750,-0.90927872)\\[1ex]\hline
\multicolumn{4}{l}{}&\multicolumn{2}{c}{$A_1=0.0015$}&\multicolumn{4}{c}{}\\[1.5ex]
{}&{}&{}&{}&{}&{}&{}&{}&{}&{}\\[-0.5ex]\hline
$B_1$ &&$L_3 (B_2=A2=M_b=0)$& $L_3 (B_2=A2=10^{-4}, M_b=0.01)$ &&$L_4 (B_2=A2=M_b=0)$& $L_4 (B_2=A2=10^{-4}, M_b=0.01)$ &&$L_5 (B_2=A2=M_b=0)$& $L_5 (B_2=A2=10^{-4}, M_b=0.01)$\\[1ex]\hline
0.0000&&(-0.19392723,-0.28949593)&(-0.20181838,-0.28650147)&&(-0.87675807,-0.82908154)&(-0.87416787,-0.82690189)&&(-0.87675800,0.82908155)&(-0.87374348,0.82648770)\\[1ex]
0.0030&&(-0.19396974,-0.28925158)&(-0.20187844,-0.28624807)&&(-0.87604795,-0.82830022)&(-0.87347669,-0.82613804)&&(-0.87823658,0.83061433)&(-0.87527601,0.82806818)\\[1ex]
0.0060&&(-0.19401316,-0.28900596)&(-0.20181838,-0.28650147)&&(-0.87534266,-0.82752417)&(-0.87416787,-0.82690189)&&(-0.87965019,0.83208249)&(-0.87374348,0.82648770)\\[1ex]
0.0090&&(-0.19405750,-0.28875905)&(-0.20200158,-0.28573736)&&(-0.87464209,-0.82675332)&(-0.87210793,-0.82462547)&&(-0.88100409,0.83349119)&(-0.87814103,0.83103030)\\[1ex]\hline
\multicolumn{4}{l}{}&\multicolumn{2}{c}{$A_1=0.0015$}&\multicolumn{4}{c}{}\\[1.5ex]
{}&{}&{}&{}&{}&{}&{}&{}&{}&{}\\[-0.5ex]\hline
$B_1$ &&$L_6 (B_2=A2=M_b=0)$& $L_6 (B_2=A2=10^{-4}, M_b=0.01)$ &&$L_7 (B_2=A2=M_b=0)$& $L_7 (B_2=A2=10^{-4}, M_b=0.01)$ &&$L_8 (B_2=A2=M_b=0)$& $L_8 (B_2=A2=10^{-4}, M_b=0.01)$\\[1ex]\hline
0.0000&&(-0.19392723,0.28949593)&(-0.20263015,0.28692262)&&(0.16829622,0.91300221)&(0.16718331,0.91029004)&&(0.16829623,-0.91300221)&(-0.20263015,0.28692262)\\[1ex]
0.0030&&(-0.19105058,0.28820483)&(-0.19971677,0.28546376)&&(0.16995578,0.91165965)&(0.16885712,0.90898002)&&(0.16772355,-0.91139458)&(0.16677409,-0.90873967)\\[1ex]
0.0060&&(-0.18832101,0.28698803)&(-0.20263015,0.28692262)&&(0.17158974,0.91032129)&(0.16718331,0.91029004)&&(0.16715340,-0.90979566)&(0.16733301,-0.91030948)\\[1ex]
0.0090&&(-0.18572192,0.28583686)&(-0.19434439,0.28279340)&&(0.17319878,0.90898719)&(0.17212759,0.90637132)&&(0.16658575,-0.90820535)&(0.16566353,-0.90562503)\\[1ex]\hline
\end{tabular}\label{tab2}}
\vspace*{-1pt}}
\end{table}

\end{landscape}
\begin{table}[!t]
{\scriptsize
\caption{\normalsize The collinear equilibrium points } \vspace*{10pt}{
\begin{tabular}{llcccccccc}\hline\hline\\[1ex]
\multicolumn{2}{l}{}&\multicolumn{3}{c}{$B_1=0.0015$, $B_2=0$, $M_b=0$ and $A_2=0$}&\multicolumn{2}{r}{}&\multicolumn{3}{r}{$B_1=0.0015$, $A_2=B_2=0.0001$, $M_b=0.01$}\\[1.5ex]
{}&{}&{}&{}&{}&{}&{}&{}&{}&{}\\[-0.5ex]\hline
$A_1$ &&&      $\omega_{1,2}[L_1]$&               $\omega_{1,2}[L_2]$&&&&                 $\omega_{1,2}[L_1]$&          $\omega_{1,2}[L_2]$\\[1ex]\hline
0.0000&&&0.00277207, -4.16745& 0.630098, -1.12595&&&&0.00629293, -4.22993&0.644868, -1.15555\\[1ex]
0.0030&&&0.00322086, -4.19389& 0.644078, -1.12368&&&&0.00693737, -4.25635&0.659084, -1.15303\\[1ex]
0.0060&&&0.00370102, -4.22042& 0.658097, -1.12159&&&&0.00761206, -4.28285&0.673337, -1.15069\\[1ex]
0.0090&&&0.00421238, -4.24702& 0.672153, -1.11969&&&&0.00831680, -4.30942&0.687623, -1.14855\\[1ex]
\hline
\multicolumn{2}{l}{}&\multicolumn{3}{c}{$A_1=0.0015$, $B_2=0$, $M_b=0$ and $A_2=0$}&\multicolumn{2}{r}{}&\multicolumn{3}{r}{$A_1=0.0015$, $A_2=B_2=0.0001$ and $M_b=0.01$}\\[1.5ex]
{}&{}&{}&{}&{}&{}&{}&{}&{}&{}\\[-0.5ex]\hline
$B_1$ &&&      $\omega_{1,2}[L_1]$&      $\omega_{1,2}[L_2]$&&&&      $\omega_{1,2}[L_1]$&      $\omega_{1,2}[L_2]$\\[1ex]\hline
0.0000&&&          -         &           -       &&&&0.01867490, -4.78680&0.649253, -1.15074\\[1ex]
0.0030&&&0.00307755, -4.20224& 0.639863, -1.12839&&&&0.00682055, -4.26512&0.654742, -1.15787\\[1ex]
0.0060&&&0.00323681, -4.24523& 0.645428, -1.13559&&&&0.00722554, -4.30892&0.660288, -1.16509\\[1ex]
0.0090&&&0.00338427, -4.28804& 0.651002, -1.14279&&&&0.00761541, -4.35251&0.665842, -1.17230\\[1ex]	  
\hline
\end{tabular}\label{tab3}}
\vspace*{-1pt}}
\end{table}
\begin{equation}
 r = \sqrt{x^2 +y^2},
\end{equation}
Furthermore, $T = a+b$, where $a$ and $b$ are two parameters which determine the density profile of the circular cluster of material points. $a$, the flatness parameter, control the  flatness of the profile while $b$, the core parameter,  controls the size of the core of the density. Now, the Lagrangian of our problem can be written as
\begin{equation}
 L=\frac{mn^2}{2}\left(x^2+y^2\right)+mn\left(x\dot{y}-\dot{x}y\right)+\frac{m}{2}\left(\dot{x}^2+\dot{y}^2\right)-V
\end{equation}
where the perturbed mean motion $n$ is define as \cite{A5}
\begin{equation}
 n^2=1+\frac{3}{2}\left(A_1+B_1-\frac{5}{4}(A_2+B_2)\right)+\frac{2M_br_c}{\left(r_c^2+T^2\right)^{3/2}}
\end{equation}
$r_c$ is the radial distance of the infinitesimal body. It then follows that the equations of motion of the infinitesimal mass are:
\begin{subequations}
\begin{equation}
\ddot{x}-2n\dot{y}=\Omega_x
\label{E9a}
\end{equation}
\begin{equation}
\ddot{y}-2n\dot{x}=\Omega_y
\label{E9b}
\end{equation}
\end{subequations}
where
\begin{equation}
\Omega=\frac{n^2(x^2+y^2)}{2}+q_1(1-2\mu)\left(\frac{1}{r_1}+\frac{A_1}{2r_1^3}-\frac{3A_2}{8r_1^5}\right)+q_2\mu \left(\frac{1}{r_2}+\frac{B_1}{2r_2^3}-\frac{3B_2}{8r_2^5}\right)+\frac{\mu}{r_3}+\frac{2M_b}{\left(r^2+T^2\right)^{1/2}}
\end{equation}
It should be noted here that the suffixes x and y indicate the partial derivatives of $\Omega$ with respect to $x$ and $y$ respectively. This system admits the well-known Jacobi integral:
\begin{equation}
C=2\Omega-(\dot{x}+\dot{y}),
\end{equation}
where C is the Jacobi constant.
\section{Equilibrium solutions}
In this section, we attempt to find the equilibrium points for the cases $y=0$ and $y\neq 0$. From the equations of motion (\ref{E9a}, \ref{E9b}), it can be deduce that equilibrium solution exists relative to the rotating frame when the partial derivatives of the pseudopotential function $\left(i.e., \frac{\partial}{\partial x}\Omega\ \ \mbox{and}\ \ \frac{\partial}{\partial y}\Omega\right)$ is equal zero. Thus, by using the following chain rule, we can find
\begin{eqnarray}
\Omega_x&=&\frac{\partial\Omega_1}{\partial x}+\frac{\partial\Omega_2}{\partial r_1}.\frac{\partial r_1}{\partial x}+\frac{\partial\Omega_3}{\partial r_2}.\frac{\partial r_2}{\partial x}+\frac{\partial\Omega_4}{\partial x}=0\nonumber\\
&&\mbox{with}\ \ \Omega_1=\frac{n^2(x^2+y^2)}{2}, \ \ \ \Omega_2=q_1(1-2\mu)\left(\frac{1}{r_1}+\frac{A_1}{2r_1^2}-\frac{3A_2}{8r_1^5}\right)\\
&&\Omega_3=q_2\mu \left(\frac{1}{r_2}+\frac{B_1}{2r_2^3}-\frac{3B_2}{8r_2^5}\right),\ \ \ \Omega_4=\frac{\mu}{r_3}+\frac{2M_b}{\left(r^2+T^2\right)^{1/2}}\nonumber
\end{eqnarray}
We have splited the differential for the sake of simplicity. Now
\begin{eqnarray}
\Omega_x&=&n^2x+\frac{q_1(2\mu-1)}{8r_1^6}\left(8r_1^4+12A_1r_1^2-15A_2\right)\left[\frac{x-\sqrt{3}\mu}{\sqrt{\left(x-\sqrt{3}\mu\right)^2+y^2}}\right]\nonumber\\
&&-\frac{\mu q_1}{8r_2^6}\left(8r_2^4+12B_1r_2^2-15B_2\right)\left[\frac{\frac{\sqrt{3}}{2}\left(1-2\mu\right)+x}{\sqrt{\left(\frac{\sqrt{3}}{2}\left(1-2\mu\right)+x\right)^2+\left(y-\frac{1}{2}\right)^2}}\right]\label{E13}\\
&&-\mu\frac{\frac{\sqrt{3}}{2}\left(1-2\mu\right)+x}{\left[\left(\frac{\sqrt{3}}{2}\left(1-2\mu\right)+x\right)^2+\left(y+\frac{1}{2}\right)^2\right]^{3/2}}-\frac{M_bx}{\left(T^2+r^2\right)^{3/2}}=0\nonumber
\end{eqnarray}
Similarly, we can find $\Omega_y$ as
\begin{eqnarray}
\Omega_y&=&n^2y+\frac{q_1(2\mu-1)}{8r_1^6}\left(8r_1^4+12A_1r_1^2-15A_2\right)\left[\frac{y}{\sqrt{\left(x-\sqrt{3}\mu\right)^2+y^2}}\right]\nonumber\\
&&-\frac{\mu q_1}{8r_2^6}\left(8r_2^4+12B_1r_2^2-15B_2\right)\left[\frac{\left(y-\frac{1}{2}\right)}{\sqrt{\left(\frac{\sqrt{3}}{2}\left(1-2\mu\right)+x\right)^2+\left(y-\frac{1}{2}\right)^2}}\right]\label{E14}\\
&&-\mu\frac{\left(y+\frac{1}{2}\right)}{\left[\left(\frac{\sqrt{3}}{2}\left(1-2\mu\right)+x\right)^2+\left(y+\frac{1}{2}\right)^2\right]^{3/2}}-\frac{M_by}{\left(T^2+r^2\right)^{3/2}}=0\nonumber
\end{eqnarray}
\begin{landscape}
\begin{table}[!t]
{\tiny
\caption{\normalsize The non-collinear equilibrium points } \vspace*{10pt}{
\begin{tabular}{ccccccccc}\hline\hline\\[1ex]
\multicolumn{3}{l}{}&\multicolumn{2}{c}{$B_1=0.0015$}&\multicolumn{4}{c}{}\\[1.5ex]
{}&{}&{}&{}&{}&{}&{}&{}&{}\\[-0.5ex]\hline
$A_1$ &$ \omega_{1,2}[L_3](B_2=A2=M_b=0)$&$\omega_{1,2}[L_3] (B_2=A2=10^{-4}, M_b=0.01)$&&$\omega_{1,2}[L_4] (B_2=A2=M_b=0)$& $\omega_{1,2}[L_4] (B_2=A2=10^{-4}, M_b=0.01)$ &&$\omega_{1,2}[L_5](B_2=A2=M_b=0)$& $\omega_{1,2}[L_5] (B_2=A2=10^{-4}, M_b=0.01)$\\[1ex]\hline
0.0000&0.698587$\pm$ 2.08943i&0.201843$\pm$ 1.95844 I&&-0.114859, -8.55824&-0.115028, -8.76845&&0.158008, -2.66934&0.146416, -2.69712\\[1ex]
0.0030&0.737719$\pm$ 2.09589i&0.241997$\pm$ 1.97939i&&-0.116121, -8.62448&-0.116285, -8.83521&&0.155672, -2.67671&0.147833, -2.70795\\[1ex]
0.0060&0.777724$\pm$ 2.10203i&0.282939$\pm$ 1.99977i&&-0.117388, -8.69093&-0.114160, -8.84750&&0.157223, -2.68772&0.149255, -2.71878\\[1ex]
0.0090&0.818571$\pm$ 2.10781i&0.324643$\pm$ 2.01957i&&-0.118659, -8.75760&-0.118812, -8.96937&&0.158780, -2.69872&0.150681, -2.72960\\[1ex]\hline	
\multicolumn{3}{l}{}&\multicolumn{2}{c}{$B_1=0.0015$}&\multicolumn{4}{c}{}\\[1.5ex]
{}&{}&{}&{}&{}&{}&{}&{}&{}\\[-0.5ex]\hline
$A_1$ &$\omega_{1,2}[L_6](B_2=A2=M_b=0)$& $\omega_{1,2}[L_6](B_2=A2=10^{-4},M_b=0.01)$ &&$\omega_{1,2}[L_7](B_2=A2=M_b=0)$& $\omega_{1,2}[L_7](B_2=A2=10^{-4},M_b=0.01)$ &&$\omega_{1,2}[L_8](B_2=A2=M_b=0)$& $\omega_{1,2}[L_8](B_2=A2=10^{-4}, M_b=0.01)$\\[1ex]\hline
0.0000&-0.112101, -2.46343&-0.139859, -2.49056&&0.243742$\pm$ 2.23115i&0.237749$\pm$ 2.26453i&&-0.082174, -1.74531&-0.160371, -1.66998\\[1ex]
0.0030&-0.128647, -2.43935&-0.155089, -2.46863&&0.247004$\pm$ 2.24285i&0.241012$\pm$ 2.27620i&&-0.066694, -1.78344&-0.142206, -1.71098\\[1ex]
0.0060&-0.145287, -2.41467&-0.170428, -2.44611&&0.250230$\pm$ 2.25450i&0.244238$\pm$ 2.28782i&&-0.051821, -1.82096&-0.124897, -1.75113\\[1ex]
0.0090&-0.162081, -2.38936&-0.185922, -2.42299&&0.253420$\pm$ 2.26611i&0.247427$\pm$ 2.29941i&&-0.037502, -1.85792&-0.108356, -1.79050\\[1ex]\hline
\multicolumn{3}{l}{}&\multicolumn{2}{c}{$A_1=0.0015$}&\multicolumn{4}{c}{}\\[1.5ex]
{}&{}&{}&{}&{}&{}&{}&{}&{}\\[-0.5ex]\hline
$B_1$ &$\omega_{1,2}[L_3](B_2=A2=M_b=0)$& $\omega_{1,2}[L_3](B_2=A2=10^{-4},M_b=0.01)$ &&$\omega_{1,2}[L_4](B_2=A2=M_b=0)$& $\omega_{1,2}[L_4](B_2=A2=10^{-4},M_b=0.01)$ &&$\omega_{1,2}[L_5](B_2=A2=M_b=0)$&$\omega_{1,2}[L_5](B_2=A2=10^{-4},M_b=0.01)$\\[1ex]\hline
0.0000&0.736505$\pm$ 2.09706i&0.240986$\pm$ 1.97780i&&-0.114430, -8.56523&-0.114588, -8.77557&&0.142775, -2.65244&0.134375, -2.68304\\[1ex]
0.0030&0.699560$\pm$ 2.08820i&0.202634$\pm$ 1.95998i&&-0.116553, -8.61745&-0.116727, -8.82805&&0.166855, -2.68976&0.159684, -2.72179\\[1ex]
0.0060&0.662538$\pm$ 2.07878i&0.224616$\pm$ 1.98507i&&-0.118691, -8.66973&-0.109272, -8.79999&&0.190282, -2.72625&0.207834, -2.80265\\[1ex]
0.0090&0.625439$\pm$ 2.06881i&0.125686$\pm$ 1.92202i&&-0.120844, -8.72209&-0.121051, -8.93323&&0.213100, -2.76199&0.208171, -2.79661\\[1ex]\hline
\multicolumn{3}{l}{}&\multicolumn{2}{c}{$A_1=0.0015$}&\multicolumn{4}{c}{}\\[1.5ex]
{}&{}&{}&{}&{}&{}&{}&{}&{}\\[-0.5ex]\hline
$B_1$ &$\omega_{1,2}[L_6](B_2=A2=M_b=0)$&$\omega_{1,2}[L_6](B_2=A2=10^{-4},M_b=0.01)$ &&$\omega_{1,2}[L_7](B_2=A2=M_b=0)$&$\omega_{1,2}[L_4](B_2=A2=10^{-4},M_b=0.01)$&&$\omega_{1,2}[L_5](B_2=A2=M_b=0)$& $\omega_{1,2}[L_5](B_2=A2=10^{-4},M_b=0.01)$\\[1ex]\hline
0.0000&-0.125093, -2.47469&-0.150461, -2.50474&&0.242486$\pm$ 2.22978i&0.236472$\pm$ 2.26310i&&-0.0667671, -1.77417&-0.150461, -2.50474\\[1ex]
0.0030&-0.114483, -2.43065&-0.143036, -2.45742&&0.248260$\pm$ 2.24422i&0.242289$\pm$ 2.27762i&&-0.0821114, -1.75454&-0.160261, -1.67929\\[1ex]
0.0060&-0.099783, -2.39535&-0.184033, -2.07113&&0.253998$\pm$ 2.25862i&0.235711$\pm$ 2.28097i&&-0.0981591, -1.73406&-0.152771, -1.70977\\[1ex]
0.0090&-0.081929, -2.36725&-0.114337, -2.39141&&0.259703$\pm$ 2.27297i&0.253809$\pm$ 2.30653i&&-0.1149660, -1.71269&-0.199014, -1.63134\\[1ex]\hline
\end{tabular}\label{tab4}}
\vspace*{-1pt}}
\end{table}
\end{landscape}
\begin{figure}[!t]
\centering\includegraphics[height=100mm,width=160mm]{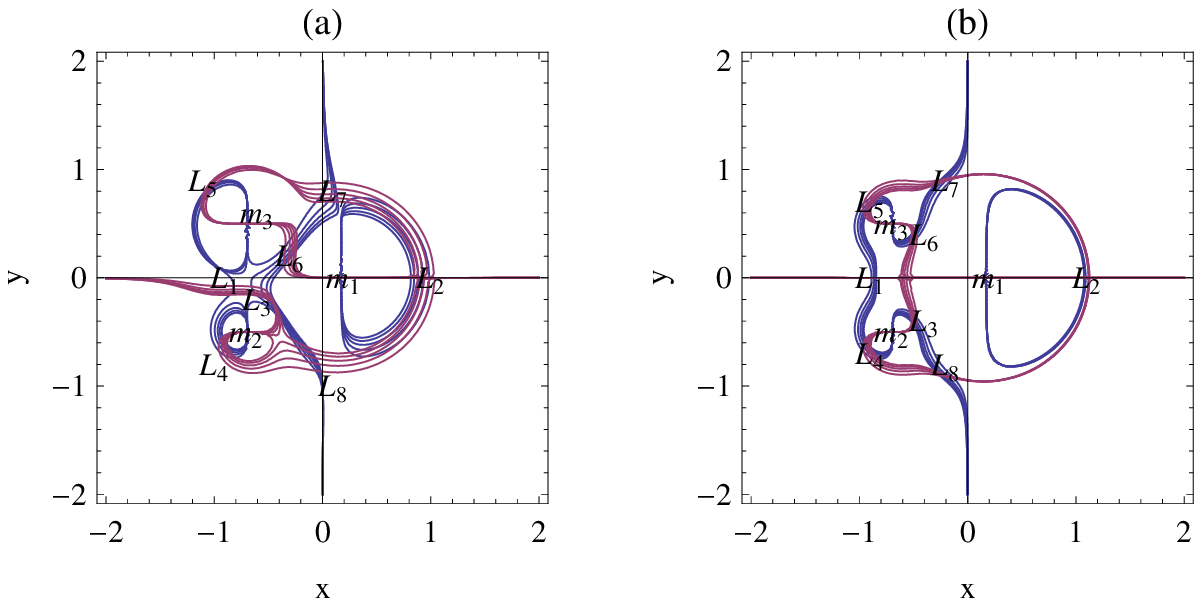}
\caption{{\protect\small (a) Graph indicating collinear and non-collinear points. The point of intersection of the two curves (blue and red) denotes the position of the equilibrium points while the double point at the center of figure-eight-shape represent the position of the primaries bodies $m_2$ and $m_3$ and $m_1$. We choose $\mu=0.2$, $q_1=q_2=1$, $r_c=0.99$, $T=0.01$, $A_1=0.0015$, $B_1=0.2:0.2:1$, $M_b=0$, $A_2=0$, $B_2=0$  (b) Same as (a) but for $B_1=0.0015$ and $A_1=0.2:0.2:1$.}}
\label{fig1}
\end{figure}
\begin{figure}[!t]
\centering\includegraphics[height=100mm,width=160mm]{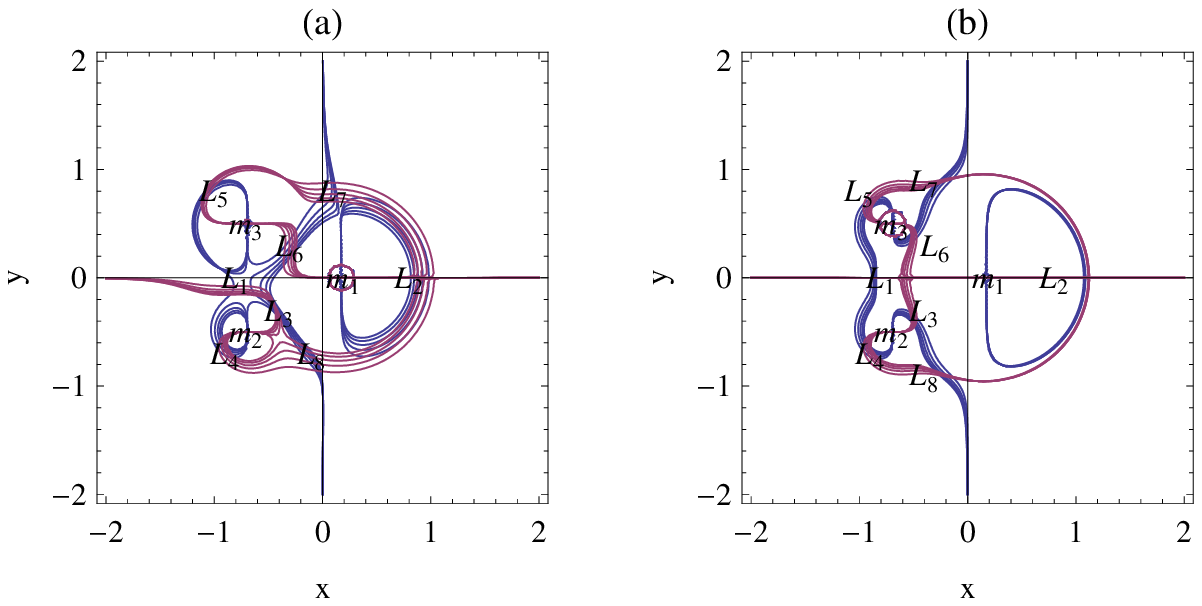}
\caption{{\protect\small (a) Graph indicating collinear and non-collinear points. The point of intersection of the two curves (blue and red) denotes the position of the equilibrium points while the double point at the center of figure-eight-shape represent the position of the primaries bodies $m_2$ and $m_3$ and $m_1$. We choose $\mu=0.2$, $q_1=q_2=1$, $r_c=0.99$, $T=0.01$, $A_1=0.0015$, $B_1=0.2:0.2:1$, $M_b=0.01$, $A_2=0.0001$, $B_2=0.0001$  (b) Same as (a) but for $B_1=0.0015$ and $A_1=0.2:0.2:1$.}}
\label{fig2}
\end{figure}
\subsection{Case 1: $y=0$ (collinear points)}
Setting y as zero, we can find the equilibrium points at x-axis by solving equations (\ref{E13}) and (\ref{E14}). This implies that collinear points lie on the line joining the primaries. We choose $\mu=0.2$ and varies the oblateness coefficients. Thus,
{\begin{eqnarray}
f(x,0)&=&n^2x+\frac{q_1(2\mu-1)}{\left(x-\sqrt{3}\mu\right)^3}\left[|x-\sqrt{3}\mu|\left(8|x-\sqrt{3}\mu|+12A_1\right)-15A_2\right]\nonumber\\
&&-\frac{\mu q_1\left(\left|\left(x+\frac{\sqrt{3}}{2}(1-2\mu)\right)^2+\frac{1}{4}\right|\left[8\left|\left(x+\frac{\sqrt{3}}{2}(1-2\mu)\right)^2+\frac{1}{4}\right|+B_1\right]-15B_2\right)}{8\left[\left(x+\frac{\sqrt{3}}{2}(1-2\mu)\right)^2+\frac{1}{4}\right]^{7/2}}\nonumber\\
&\times&\left(\frac{\sqrt{3}}{2}\left(1-2\mu\right)+x\right)-\mu\frac{\frac{\sqrt{3}}{2}\left(1-2\mu\right)+x}{\left[\left(\frac{\sqrt{3}}{2}\left(1-2\mu\right)+x\right)^2+\frac{1}{4}\right]^{3/2}}-\frac{M_bx}{\left(T^2+r^2\right)^{3/2}}=0\label{E15}
\end{eqnarray}}
\begin{eqnarray}
g(x,0)&=&\frac{\mu q_1\left(\left|\left(x+\frac{\sqrt{3}}{2}(1-2\mu)\right)^2+\frac{1}{4}\right|\left[8\left|\left(x+\frac{\sqrt{3}}{2}(1-2\mu)\right)^2+\frac{1}{4}\right|+B_1\right]-15B_2\right)}{16\left[\left(x+\frac{\sqrt{3}}{2}(1-2\mu)\right)^2+\frac{1}{4}\right]^{7/2}}\nonumber\\
&&-\mu\frac{1}{\left[\left(\frac{\sqrt{3}}{2}\left(1-2\mu\right)+x\right)^2+\frac{1}{4}\right]^{3/2}}=0\label{E16}
\end{eqnarray}
The two collinear points have been represented in figures (\ref{fig1}-\ref{fig2}) and presented for better clarification in table (\ref{tab1})
\subsection{Case 2: $y\neq0$ (non-collinear points)}
The non-collinear points can be found by solving equations (\ref{E13}) and (\ref{E14}) simultaneously when $y\neq0$, i.e. $f(x,y)=0$ and $g(x,y)=0$, where
\begin{subequations}
\begin{eqnarray}
f(x,y)&=&n^2x+\frac{q_1(2\mu-1)}{8r_1^6}\left(8r_1^4+12A_1r_1^2-15A_2\right)\left[\frac{x-\sqrt{3}\mu}{\sqrt{\left(x-\sqrt{3}\mu\right)^2+y^2}}\right]\nonumber\\
&&-\frac{\mu q_1}{8r_2^6}\left(8r_2^4+12B_1r_2^2-15B_2\right)\left[\frac{\frac{\sqrt{3}}{2}\left(1-2\mu\right)+x}{\sqrt{\left(\frac{\sqrt{3}}{2}\left(1-2\mu\right)+x\right)^2+\left(y-\frac{1}{2}\right)^2}}\right]\label{E17a}\\
&&-\mu\frac{\frac{\sqrt{3}}{2}\left(1-2\mu\right)+x}{\left[\left(\frac{\sqrt{3}}{2}\left(1-2\mu\right)+x\right)^2+\left(y+\frac{1}{2}\right)^2\right]^{3/2}}-\frac{M_bx}{\left(T^2+r^2\right)^{3/2}}=0\nonumber
\end{eqnarray}
\begin{eqnarray}
g(x,y)&=&n^2y+\frac{q_1(2\mu-1)}{8r_1^6}\left(8r_1^4+12A_1r_1^2-15A_2\right)\left[\frac{y}{\sqrt{\left(x-\sqrt{3}\mu\right)^2+y^2}}\right]\nonumber\\
&&-\frac{\mu q_1}{8r_2^6}\left(8r_2^4+12B_1r_2^2-15B_2\right)\left[\frac{\left(y-\frac{1}{2}\right)}{\sqrt{\left(\frac{\sqrt{3}}{2}\left(1-2\mu\right)+x\right)^2+\left(y-\frac{1}{2}\right)^2}}\right]\label{E17b}\\
&&-\mu\frac{\left(y+\frac{1}{2}\right)}{\left[\left(\frac{\sqrt{3}}{2}\left(1-2\mu\right)+x\right)^2+\left(y+\frac{1}{2}\right)^2\right]^{3/2}}-\frac{M_by}{\left(T^2+r^2\right)^{3/2}}=0\nonumber
\end{eqnarray}
\end{subequations}
We have also choose $\mu=0.2$ and varies the oblateness coefficients. The six non-collinear points have been represented in figures (\ref{fig1}-\ref{fig2}) and presented for better clarification in table (\ref{tab2})
\section{Stability of libration points}
In this section, we aim to study the stability of a libration point $(x_0,y_0)$. To achieve this goal, firstly, we apply infinitesimal displacement $\zeta$ and $\eta$ to the coordinates via:
\begin{equation}
\eta=y-y_0\ \ \ \ \ \zeta=x-x_0
\label{E18}
\end{equation}
Substituting equation (\ref{E18}) into equations of the motion in (\ref{E9a}) and  (\ref{E9b}), we can find
\begin{equation}
\ddot{\zeta}-2n\dot{\eta}=\zeta\Omega_{xx}^0+\eta\Omega_{xy}^0\ \ \ \ \ \ \ddot{\eta}+2n\dot{\zeta}=\zeta\Omega_{yx}^0+\eta\Omega_{yy}^0,
\label{E19}
\end{equation}
where the superfix `0' indicates that the partial derivatives have been computed at the triangular libration point by considering $(x_0,y_0)$. Now, let us assume a solution of the form $\zeta=C_1\exp^{\lambda t}$ and $\eta=C_2\exp^{\lambda t}$, where $C_1$ and $C_2$ is a constant and $\lambda$ is a parameter. Substituting the assumed solutions in equation (\ref{E19}), we can find the following non-trivial solution for $C_1$ and $C_2$
\begin{equation}
\left|\begin{matrix}\lambda^2-\Omega_{xx}^0&&-2n\lambda-\Omega_{xy}^0\\ 2n\lambda-\Omega_{xy}^0&&\lambda^2-\Omega_{yy}^0 \end{matrix}\right|=0
\label{E20}
\end{equation}
Solving the determinant by expansion, we can find the characteristic equation corresponding to the variational equations of (\ref{E19}) as
\begin{equation}
\lambda^4+(4n^2-\Omega_{xx}^0-\Omega_{yy}^0)\lambda^2+\left(\Omega_{xx}^0\Omega_{yy}^0-{\Omega_{xy}^0}^2\right)=0
\label{E21}
\end{equation}
Now, the second order derivatives of the potential function w.r.t x and y are as follows
\begin{subequations}
\begin{eqnarray}
\Omega_{xx}&=&\frac{(2 \mu-1) {q}_1 \left[3 y^2 \left(20 {A}_1 {r}_1^2-35 {A}_2+8 {r1}^4\right)-16 \left(3 {A}_1 {r}_1^4+{r}_1^6\right)+90 {A}_2 {r}_1^2\right]}{8 {r}_1^9}\nonumber\\
&&+\frac{\mu {q}_2 \left[4 {r}_2^2 \left(3 {B}_1 \left(16 {r}_2^2-5 (1-2 y)^2\right)+2 {r}_2^2 \left(8 {r}_2^2-3 (1-2 y)^2\right)\right)-15 {B}_2 \left(24 {r}_2^2-7 (1-2 y)^2\right)\right]}{32 {r}_2^9}\nonumber\\
&&+\frac{\mu \left(8 {r}_3^2-3 (2 y+1)^2\right)}{4{r}_3^5 }+n^2-\frac{{M}_b \left(T^2-2 x^2+y^2\right)}{\left(T^2+x^2+y^2\right)^{5/2}}
\label{E21a}
\end{eqnarray}
\begin{eqnarray}
\Omega_{yy}&=&\frac{q_1(2\mu-1)\left[(x-\sqrt{3}\mu)^2\left(24r_1^4+60r_1^2A_1-105A_2\right)-16r_1^6-48r_1^4A_1+90r_1^2A_2\right]}{8r_1^9}\nonumber\\
&&-\frac{\mu q_2(3B_1+4r_2^2)\left(y-\frac{1}{2}\right)^2}{r_2^7}+\frac{7\mu q_2\left(-15B_2+12B_1r_2^2+8r_2^4\right)\left(y-\frac{1}{2}\right)^2}{8r_2^9}\label{E21b}\\
&&-\frac{\mu}{r_3^3}+\frac{3\mu\left(y+\frac{1}{2}\right)^2}{r_3^5}+n^2-\frac{M_b(T^2+x^2-2y^2)}{(T^2+x^2+y^2)^{5/2}}-\frac{\mu q_2\left(-15B_2+12B_1r_2^2+8r_2^4\right)}{8r_2^7}\nonumber
\end{eqnarray}
\begin{eqnarray}
\Omega_{xy}=-\Omega_{yx}&=&\frac{3\mu q_2\left(35B_2+20B_1r_2^2+8r_2^4\right)\left(2\sqrt{3}\mu-\sqrt{3}-2x\right)\left(y-\frac{1}{2}\right)}{16r_2^9}\nonumber\\
&&+\frac{3(2\mu-1)q_1\left(8r_1^4+20A_1r_1^2-35A2\right)\left(\sqrt{3}\mu-x\right)y}{8r_1^9}\label{E21c}\\
&&+\frac{3M_bxy}{(T^2+x^2+y^2)^{5/2}}+\frac{3\mu\left(x+\frac{\sqrt{3}}{2}(1-2\mu)\right)\left(y+\frac{1}{2}\right)}{r_3^5}\nonumber
\end{eqnarray}
\end{subequations}
\subsection{Stability of collinear points}
For stability of the collinear points $(x_0, 0)$, we have
\begin{subequations}
\begin{eqnarray}
\Omega_{xx}^{0}&=&\frac{(2 \mu-1) {q}_1 \left[90 {A}_2 r_{*1}^2-16 \left(3 {A}_1 r_{*1}^4+r_{*1}^6\right)\right]}{8r_{*1}^9}+\frac{\mu \left(8 r_{*3}^2-3\right)}{4r_{*3}^5 }+n^2-\frac{{M}_b \left(T^2-2x^2\right)}{\left(T^2+x^2\right)^{5/2}}\nonumber\\
&&+\frac{\mu {q}_2 \left[4 r_{*2}^2 \left(3 {B}_1 \left(16 r_{*2}^2-5 \right)+2 r_{*2}^2 \left(8 r_{*2}^2-3\right)\right)-15 {B}_2 \left(24 r_{*2}^2-7\right)\right]}{32 r_{*2}^9}
\label{E22a}
\end{eqnarray}
\begin{eqnarray}
\Omega_{yy}^{0}&=&\frac{q_1(2\mu-1)\left[(x-\sqrt{3}\mu)^2\left(24r_{*1}^4+60r_{*1}^2A_1-105A_2\right)-16r_{*1}^6-48r_{*1}^4A_1+90r_{*1}^2A_2\right]}{8r_{*1}^9}\nonumber\\
&&-\frac{\mu q_2(3B_1+4r_{*2}^2)}{4r_{*2}^7}+\frac{7\mu q_2\left(-15B_2+12B_1r_{*2}^2+8r_{*2}^4\right)}{32r_{*2}^9}\label{E22b}\\
&&-\frac{\mu}{r_{*3}^3}+\frac{3\mu}{4r_{*3}^5}+n^2-\frac{M_b}{(T^2+x^2)^{3/2}}-\frac{\mu q_2\left(-15B_2+12B_1r_{*2}^2+8r_{*2}^4\right)}{8r_{*2}^7}\nonumber
\end{eqnarray}
\begin{eqnarray}
\Omega_{xy}^{0}=\Omega_{yx}=\frac{3\mu q_2\left(35B_2+20B_1r_{*2}^2+8r_{*2}^4\right)\left(2\sqrt{3}\mu-\sqrt{3}-2x\right)}{32r_{*2}^9}+\frac{3\mu\left(x+\frac{\sqrt{3}}{2}(1-2\mu)\right)}{2r_{*3}^5},
\label{E22}
\end{eqnarray}
\end{subequations}
where we have introduced parameters $r_{*1}=\left|x-\sqrt{3}\mu\right|, r_{*2}=r_{*3}=\sqrt{\left(x+\frac{\sqrt{3}}{2}(1-2\mu)\right)^2+\frac{1}{4}}$ for mathematical simplicity. Substituting these values given by equation (22) into the characteristic equation, we have
\begin{eqnarray}
\lambda_1=-\lambda_2=\sqrt{\frac{\Omega_{xx}^0+\Omega_{yy}^0-4n^2+\sqrt{(\Omega_{xx}^0+\Omega_{yy}^0-4n^2)^2-4\Omega_{xx}^0\Omega_{yy}^0}}{2}}\\
\lambda_3=-\lambda_4=\sqrt{\frac{\Omega_{xx}^0+\Omega_{yy}^0-4n^2-\sqrt{(\Omega_{xx}^0+\Omega_{yy}^0-4n^2)^2-4\Omega_{xx}^0\Omega_{yy}^0}}{2}}
\end{eqnarray}
Numerical computation of $\lambda_i (i=1,2,3,4)$ have been presented in tables \ref{tab3}. We take $\lambda_{1,2}^2=\omega_1$ and $\lambda_{3,4}^2=\omega_2$. It can be seen that collinear equilibrium points are unstable for all the variational groups we considered.
\subsection{Stability of non-collinear points}
Again, we solve the characteristic equation for $\lambda_i (i=1,2,3,4)$, i.e. 
\begin{eqnarray}
\lambda_1=-\lambda_2=\sqrt{\frac{\Omega_{xx}^0+\Omega_{yy}^0-4n^2+\sqrt{(\Omega_{xx}^0+\Omega_{yy}^0-4n^2)^2-4\Omega_{xx}^0\Omega_{yy}^0}}{2}}\\
\lambda_3=-\lambda_4=\sqrt{\frac{\Omega_{xx}^0+\Omega_{yy}^0-4n^2-\sqrt{(\Omega_{xx}^0+\Omega_{yy}^0-4n^2)^2-4\Omega_{xx}^0\Omega_{yy}^0}}{2}}
\end{eqnarray}
and then evaluate them at equilibrium points. The results have been presented in table \ref{tab4}. We take $\lambda_{1,2}^2=\omega_1$ and $\lambda_{3,4}^2=\omega_2$. An equilibrium point is stable if the characteristic equation evaluated at the equilibrium point, has four complex roots with negative real parts or pure imaginary roots otherwise it is unstable. In this regard, the non-collinear equilibrium points are found to be stable. 

By taking $\mu=0$ and using $L_4(-0.87342321, -0.82624651)$, we have the discriminant $\Delta=(\Omega_{xx}^0+\Omega_{yy}^0-4n^2)^2-4\Omega_{xx}^0\Omega_{yy}^0= 4.69628$ $(\Delta>0)$ and by taking $\mu=\frac{1}{2}$ with $L_4(-0.87342321, -0.82624651)$, we have the discriminant $\Delta=(\Omega_{xx}^0+\Omega_{yy}^0-4n^2)^2-4\Omega_{xx}^0\Omega_{yy}^0= -0.5180488$ $(\Delta<0)$. This implies that $\Delta$ is a decreasing function of $\mu$ within an interval $\left(0,\frac{1}{2}\right)$ and there exist only one value of $\mu$ called as critical mass ($\mu_c$) in the interval $\left(0,\frac{1}{2}\right)$ for which the discriminant is zero.
\section{Results and Conclusion}
In this research, we examined the motion of an infinitesimal mass by assuming that the primaries of the system are radiating-oblate spheroids surrounded by a circular cluster of material points, within the framework of restricted four-body problem. In our model we assume that the two masses of the primaries $m_2$ and $m_3$ are equal to $\mu$ and the mass $m_1$
is $1-2\mu$. we have obtained the equilibrium points via numerical computation. In table \ref{tab1} we obtained the two collinear points $L_{1}$ and $L_2$ on the x-axis. Firstly, we fixed $B_1$ at 0.0015 in the absent of gravitational potential with $A_2=B_2=0$ and then varies $A_1$ as 0.0000:0.0030:0.0090. We found that the equilibrium point shifted from left to right. This is also observed in the presence of gravitational potential with $A_2=B_2\neq 0$. However, when we fixed $A_1$ at 0.0015 and then varies $B_1$, we found that collinear point $L_1$ shifted from left to right whereas that of $L_2$ shifted from right to left. This is observed in the two cases we considered, i.e. $A_2=B_2=M_b=0$ and $A_2=B_2\neq M_B\neq 0$. 

In table \ref{tab2}, we present the non-collinear points for the two cases  $A_2=B_2=M_b=0$ and $A_2=B_2\neq M_B\neq 0$. It is shown that $L_{4}(x, y)\approx L_{5}(x, -y)$. We found that our results are in excellent agreement with Kumari and Kushvah \cite{N2} when $A_2=B_2=M_b=0$ also with the ones obtained by Papadouris and Papadakis \cite{N3} when  $A_1=B_1=A_2=B_2=M_b=0$. We also examined the linear stability of these point and found then to be unstable. An extension to the effect of potential created by the circular cluster and oblateness coefficients for the more massive primary and the less massive primary, on the existence and linear stability of the libration point have also been presented. 

libration points are very important in astronomy because they indicate places where particle can be trapped. As a consequence, our results can be applied in astrophysics


\begin{thebibliography}{99}
\bibitem{N1} J. Singh and J. J. Taura, Astrophys Space Sci {\bf350} (2014) 127.
\bibitem{A2} J. Singh and J. J. Taura, Astrophys Space Sci {\bf351} (2014) 499.
\bibitem{A1} A. N. Baltagiannis and K. E. Papadakis,  Astrophys Space Sci {\bf336} (2011) 357.
\bibitem{N2} R. Kummari and B. S. Kushvah,  Astrophys. Space Sci. {\bf349} (2014) 693.
\bibitem{N3} J. P. Papadouris and K. E. Papadakis, Astrophys. Space Sci. {\bf344} (2013) 21.
\bibitem{N4} M. Alvarez and C. Vidal, Math. Prob. Eng, doi:10.1155/2009/181360.
\bibitem{A3} V. V. Radzievskii,  Astron. J. {\bf27} (1950) 250.
\bibitem{A4} M. Miyamoto, R. Nagai, Publ. Astron. Soc. Jpn. 27, 533 (1975)
\bibitem{A5} I. D. Peter, J. J. Lissauer, Planetary Science. Cambridge University Press, New York (2001)
\end{thebibliography}
\end{document}